\documentclass[12pt]{article}
\usepackage[english]{babel}
\usepackage{graphicx}
\usepackage{amsmath}

\large \baselineskip = 20 true pt

\begin{document}
\begin{center}
{\Large \bf Severity Level of Permissions in Role-Based Access Control} \vspace{0.5cm}
\end{center}

\begin{center}
S.V. Belim,  N.F. Bogachenko, A.N. Kabanov\\
Dostoevsky Omsk State University, Omsk, Russia

 \vspace{0.5cm}
\end{center}

\begin{center}
{\bf Abstract}
\end{center}

The analysis of hidden channels of information leakage with respect to role-based access control includes monitoring of excessive permissions among users. It is not always possible to completely eliminate redundancy. The problem of ranking permissions arises in order to identify the most significant, for which redundancy is most not desirable. A numerical characteristic that reflects the value or importance of permissions is called the "severity level". A number of heuristic assumptions have been formulated that make it possible to establish the dependence of the severity level of permissions on the structure of the role hierarchy. A methodology for solving the problem is proposed, using analytic hierarchy process and taking into account these assumptions. The main idea is that the decision tree of the process will be the role graph.


\section{Introduction}

The concept of roles has become widespread in the field of information security~\cite{b1, b2}. At the heart of the role-based access control is the idea of distributing rights and privileges (we will call them permissions) between users not directly, but through assigning certain roles to them. In the process of authorizing a user for a certain role, he receives a set of permissions assigned to the given role.

One of the most difficult problems encountered in the construction of access control systems is the problem of covert channels of information leakage. A covert channel of information leakage is a mechanism by which a computer system can transmit information between entities bypassing the access control policy. At the same time, any model of access control should provide proof, including formal one, of the impossibility of inadmissible information flows. But such proofs are mostly based on a single security criterion. For example, such a criterion may be "the impossibility of accessing subjects to objects outside explicit permissions". Outside the proof, there are implicit information flows, which are caused by covert channels of information leakage. Thus, proof of the absence of "dangerous" accesses and analysis of covert channels of information leakage are two components of any access control system. 

In the context of a role-based access control, for which many modifications and extensions have been proposed~\cite{b3,b4,b5,b6,b7,b8,b9,b10}, first of all an analysis of the problem of users' excessive permissions is needed. Obviously, this indicator is decisive in assessing the possibility of unauthorized transmission of information. As part of the solution of the problem, permissions should be ranked "according to the level of danger of information leakage" or "according to the degree of preference from the point of view of the attacker". The quantitative characteristics that meet the qualitative criteria are called "the severity level" of permissions. The term used is understood as the significance or value of the corresponding rights and privileges in terms of information security of the system: the higher the severity level of permission, the greater the loss in case of its misuse (leakage).

In this paper, we consider systems with a hierarchical organization of a set of roles that contain in their description an oriented (role) graph, which is most often used only to control the inheritance of the permissions assigned to roles. It is suggested, on the basis of the role graph, to obtain additional information necessary for analyzing the security of the system. Most approaches to analyzing graphs of access to information resources are connected with the study of attack graphs of models with discretionary access control~\cite{b11}. In a more general case, the threats realization graph can be constructed to analyze the security of information systems~\cite{b12}. Unfortunately, most of the work is devoted to the synthesis of such graphs and only a few to formal models of their analysis. Most often this is the use of standard methods of analysis of oriented graphs~\cite{b13}, or other search algorithms, for example, recursive algebraic analysis~\cite{b14}. As will be shown below, the hierarchical organization of the set of roles in a system with role-based access control makes it possible to apply the analytic hierarchy process, used to quantify the phenomena of a qualitative nature, to determine the severity level and rank permissions~\cite{b15, b16}. In the future, when reconstructing the role hierarchy in order to reduce the number of excessive permissions, it is first necessary to analyze roles that have permissions with a higher severity level.

\section{Formulation of the problem}

The role-based access control is determined by a set of four sets: $U$ -- the set of users, $R$ -- the set of roles, $P$ -- the set of permissions (rights, privileges), $C$ -- the set of users' sessions in the system. Along with the role, the user receives a certain set (list) of permissions assigned to the given role. There are three main issues in the organization of role-based access:

1. How many and which roles can be assigned to one user.  

2. How many and which roles a user can use in a single session with the system (to which roles the user can be authorized).

3. Is it possible to delegate (transfer) permissions from one role to another.

Depending on the methods of solving the questions posed, there are several types of role models. The most common is a model with a hierarchical organization of multiple roles. In this case, the senior in the hierarchy role gets the permissions of the roles directly subordinate to it.

We denote the set of permissions assigned to some role $r \in R$ as $RP(r)$. Then the model with a hierarchical role system is determined by the following refinements and additions:

1. A partial order relation is introduced on the role set $R$ specifying the dominant / subordinate operator of the roles "$\geq$": if $r_1 \geq r_2$, then the role $r_1$ is higher in the hierarchy than the role $r_2$ and the set of permissions $RP(r_1)$ includes the set of permissions $RP(r_2)$.

2. If the role $r$ is assigned to the user $u \in U$, then all roles subordinate to $r$ are automatically assigned to him. If the user $u$ is authorized for the role $r$ in the current session of the system, then he is automatically authorized for all roles that are subordinate to $r$. The set of permissions that are available to the user for all roles for which he is authorized in the current session of the system is determined by taking into account the subordinate roles.

It is convenient to describe the hierarchy of roles with a labeled oriented graph (we call it role), in which the vertices correspond to the roles existing in the system, the vertex labels represent the sets of permissions for these roles, the edges define the dominance / subordination of roles. As a rule, it is considered that the role graph is an oriented tree. In fact, this is not so obvious, and in a more general case it is sufficient to require that the hierarchy be described by an oriented acyclic graph~\cite{b17}. But within the framework of this article, we need the tree role graph. It is worth noting that the authors of~\cite{b9, b17} studied the problem of transforming an arbitrary oriented graph without cycles describing the hierarchy of roles into an equivalent tree and presented algorithms for its solution.

So, let the hierarchy of roles be given by the oriented tree $T$ -- the role tree. At this stage of the problem analysis, we assume that in the role hierarchy, only leaf nodes directly receive the permissions, and then the permissions are distributed according to the principle of inheritance (using the leaf approach to the distribution of permissions~\cite{b17}).

By the severity level of the permission $p_i \in P$ we call the numerical characteristic $S(p_i)$ characterizing the significance or value of $p_i$ from the point of view of system's information security. We require that $\forall p_i \in P$: $0 \leq S(p_i) \leq 1$, $\sum S(p_i) = 1$ (summation is over all $i$ from 1 to $m$, where $m$ is the number of permissions). 

Obviously, the value of $S(p_i)$ should depend both on the prevalence of the permission $p_i$ in the system, and on the significance of the roles to which this permission is assigned (that is, on the hierarchical organization of the set $R$). The task is set for each element $p_i$ of the permission set $P$ to calculate its severity level $S(p_i)$. We will proceed from the following heuristic assumptions.

\textit{Assumption 1.} The more permissions this role contains, the more likely it will be attacked, the greater is the severity level of permissions assigned to the role. 

\textit{Assumption 2.} The more often this permission appears in the list of roles permissions, the more likely it is to leak, the greater is its severity level.

\textit{Assumption 3.} The higher the role in the hierarchy, the more likely it will be attacked, the greater is the severity level of permissions assigned to the role.

\section{Hierarchy of roles and analytic hierarchy process}

Let's estimate the severity level of permissions using the analytic hierarchy process (AHP), well-known from decision support theory~\cite{b15, b16}. The main difficulty of AHP is the search for quantitative indicators for constructing a numerical scale of preferences for possible alternatives. Often, a mechanism of expert assessments is used to do this. In turn, subjectivism and inconsistency of expert judgments are the main sources of criticism of the method~\cite{b18}. In this paper, we propose to use the hierarchical organization of the set of roles to construct the AHP decision tree, and calculate the coefficients of the pairwise comparison matrices based on the distribution of permissions in the system, without involving external subjective expert assessments.

The proposed approach consists of several stages. The first one preprocesses the role hierarchy: each leaf node $r_l$ of the original role tree $T$ is appended with additional vertices, each of which contains exactly one permission from the set of permissions $RP(r_l)$ of the vertex $r_l$. Thus, an extended role tree $T_p$ is constructed. The decision tree (or hierarchy) of the AHP will be the tree $T_p$.

The second stage calculates the relative weights of all vertices (except the root) of the tree $T_p$. Calculation of weights occurs when moving from the root to the leaf vertices. At each step, a subset of the roles $\{r_1, \ldots, r_k\}$, subordinate to one role $r$ of the previous level, is considered. In the AHP terminology, the selected vertices are the matched factors or alternatives for the $r$ criterion. For each such subset a pairwise comparison matrix $\textbf{M}$ is constructed. The dimension $k$ of the matrix $\textbf{M}$ is equal to the cardinality of the allocated subset of the subordinate roles. The element $m_{ij} = [\textbf{M}]_{ij}$ in the pairwise comparison matrix corresponding to the pair $(r_i, r_j)$ is equal to the ratio of the number of permissions $|RP(r_i)|$ of role $r_i$ to the number of permissions $|RP(r_j)|$ of role $r_j$:
$$m_{ij} = \frac{|RP(r_i)|}{|RP(r_j)|}.$$

Taking into account Assumption 1, the value $m_{ij}$ characterizes the degree of significance of the role $r_i$ in comparison with the role $r_j$ from the point of view of the severity level of permissions. The diagonal elements $m_{ii}$ are equal to one. Then the columns of the matrix $\textbf{M}$ are normalized, and the elements $m^*_{ij}$ of the new normalized matrix $\textbf{M}^*$ are calculated by the formula: $m^*_{ij} = m_{ij} / (m_{1j} + \ldots + m_{kj})$. The weight of each role $r_i$ is calculated as the arithmetic mean of the corresponding row role in the normalized matrix $\textbf{M}^*$: $w_i = (m^*_{i1} + \ldots + m^*_{ik}) / k$. 

Since the elements $m_{ij}$ of the pairwise comparison matrix are calculated automatically, the ideal consistency of the matrix $\textbf{M}$ is ensured: its elements are connected by the equalities $m_{ij} = m_{is} \times m_{sj}$ (for any $i$, $j$, $s$). Indeed,
$$m_{is} \times m_{sj} = \frac{|RP(r_i)|}{|RP(r_s)|} \times \frac{|RP(r_s)|}{|RP(r_j)|} = \frac{|RP(r_i)|}{|RP(r_j)|} = m_{ij}.$$
Since the matrix $\textbf{M}$ is ideal consistent, after normalization the columns of the matrix $\textbf{M}^*$ become the same. Hence, $w_i = m^*_{ij}$ for any $j$. Without loss of generality, we will continue to work with the elements of the first column of the matrix $\textbf{M}$, then
$$w_i = m^*_{i1} = \frac{m_{i1}}{m_{11} + \ldots + m_{k1}} = \frac{|RP(r_i)|}{|RP(r_1)| + \ldots + |RP(r_k)|},$$ 
for $i = 1, \ldots, k$. Thus, the relative weight coefficients can be calculated without constructing a pairwise comparison matrix, only on the basis of knowledge of some numerical characteristics attributed to the compared factors (roles). In our case, this is the number of permissions. Thus, in the problem under consideration, AHP is based only on the properties of the system itself, which frees it from the "model" error that arises from the inconsistency of expert assessments~\cite{b18}.

At the third stage, the combined weights of the leaf vertices of the tree $T_p$ are calculated. These values will be the severity level of permissions. The severity level $S(p)$ of each permission $p$ is equal to the sum of the products of the relative weight coefficients of the vertices along all the paths from the root to the leaves containing this permission. Note that $\forall i = 1, \ldots, |R|$: $0 \leq w_i \leq 1$. Therefore, at this stage, Assumption~2 is taken into account: the more items, the greater is $S(p)$, and Assumption~3: the shorter the path, the larger is product, the larger is the contribution to $S(p)$.

\section{Algorithm for calculating the severity level of permissions}

Let's write out formally the algorithm for calculating the severity level of permissions. Let use the leaf principle of distribution of permissions; the role tree $T$, which defines the hierarchy of roles, contains $n$ vertices $r_1, \ldots, r_n$; $m$ permissions $p_1, \ldots, p_m$ are defined in the system. The algorithm for steps can be written in the following form.

1. To each vertex $r_i$ of the role tree $T$, assign the value $|RP(r_i)|$ equal to the number of permissions assigned to this vertex.

2. Construct an extended role tree $T_p$: add as many permission vertices to each leaf node $r_l$ of the role tree $T$, as contained in its permission set $RP(r_l)$; assign to each of the new vertices one permission from the set of permissions $RP(r_l)$ and assign to it a numerical characteristic (see step~1) equal to one.

3. For each non-leaf vertex $r_i$ of the tree $T_p$, starting from the root, consider a subset of the vertices subordinate to it: $\{r_{i1}, \ldots, r_{ik}\}$. For the vertices of this subset, calculate the relative weights by the formula: 
$$w_{ij} = \frac{|RP(r_{ij})|}{|RP(r_{i1})| + \ldots + |RP(r_{ik})|},\ j = 1, \ldots, k.$$ 	 

4. For each permission $p_i$, $i = 1, \ldots, m$, calculate its severity level:
$$S(p_i)=\sum_{\rho(r,r_l):\ RP(r_l)=\{p_i\}}\left(\prod_{j:r_j\in\rho(r,r_l)}w_j \right) .$$
The sum is taken over all oriented paths $\rho(r, r_l)$ in the tree $T_p$, leading from the root $r$ to that leaf $r_l$, the list of permissions $RP(r_l)$ of which contains the single permission $p_i$ (obviously, these are the permission vertices that were added to the original role tree $T$ and which correspond to the permission $p_i$). In each product, there are the relative weight coefficients $w_j$ from step~3 of those vertices $r_j$ that compose the oriented path $\rho(r, r_l)$ (excluding the root $r$, because the weight is not defined for it). 

Note that $\forall p_i \in P:$ $0 \leq S(p_i) \leq 1$ è $\sum S(p_i) = 1$, $i = 1, \ldots, m$. Thus, the numerical characteristics found are relative values and allow a comparative analysis of the permissions.

It is easy to understand that the computational complexity of the presented algorithm depends polynomially on the number of roles $n$ and the number of permissions $m$ in the original role tree $T$ and does not exceed O$((n \cdot m)^2)$. 

\section{Example}

\begin{figure}[ht]
\centering
\includegraphics[width=0.4\textwidth]{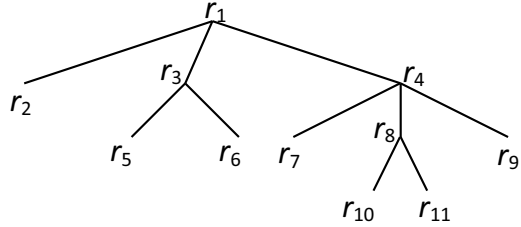}
\caption{The role tree $T$}
\label{fig_1}
\end{figure}

Consider the hierarchy of roles represented by the unbalanced tree T (see fig.~\ref{fig_1}).

\begin{figure}[ht]
\centering
\includegraphics[width=0.4\textwidth]{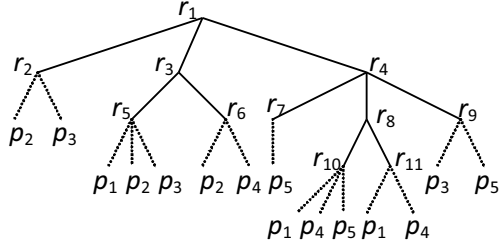}
\caption{The expanded role tree $T_p$}
\label{fig_2}
\end{figure}

Let the permissions of leaf roles are represented by the following sets: $RP(r_2) = \{p_2, p_3\}$, $RP(r_5) = \{p_1, p_2, p_3\}$, $RP(r_6) = \{p_2, p_4\}$, $RP(r_7) = \{p_5\}$, $RP(r_10) = \{p_1, p_4, p_5\}$, $RP(r_{11}) = \{p_1, p_4\}$, $RP(r_9) = \{p_3, p_5\}$, and for other roles, the permissions are determined from the inheritance condition: $RP(r_8) = \{p_1, p_4, p_5\}$, $RP(r_3) = \{p_1, p_2, p_3, p_4\}$, $RP(r_4) = \{p_1, p_3, p_4, p_5\}$, $RP(r_1) = \{p_1, p_2, p_3, p_4, p_5\}.$ The expanded $T_p$ role tree is shown in fig.~\ref{fig_2}. 

For this example $|RP(r_2)| = 2$, $|RP(r_3)| = |RP(r_4)| = 4$. Hence, $w_2 = 2 / (2 + 4 + 4) = 0.2$, $w_3 = w_4 = 4 / (2 + 4 + 4) = 0.4$. In vector form:

$$
|\textbf{RP}|_{2,3,4} = 
\begin{pmatrix} 
2 \\ 
4 \\
4
\end{pmatrix},\ 
\textbf{W}_{2,3,4} = 
\begin{pmatrix} 
0.2 \\ 
0.4 \\
0.4 
\end{pmatrix}.
$$

The coordinates of the vector $|\textbf{RP}|_{2,3,4}$ are determined by the number of permissions assigned to the roles $r_2$, $r_3$, $r_4$; the coordinates of the vector $\textbf{W}_{2,3,4}$ are determined by the relative weight coefficients of these roles.

Similarly, the coordinates of the remaining vectors are calculated:
$$
|\textbf{RP}|_{5,6} = \begin{pmatrix} 
3 \\
2
\end{pmatrix},\ 
|\textbf{RP}|_{7,8,9} = \begin{pmatrix} 
1 \\ 
3 \\
2
\end{pmatrix},\ 
|\textbf{RP}|_{10,11} = \begin{pmatrix} 
3 \\
2
\end{pmatrix},
$$
$$
\textbf{W}_{5,6} = \begin{pmatrix} 
0.6 \\
0.4
\end{pmatrix},\ 
\textbf{W}_{7,8,9} \approx \begin{pmatrix} 
0.17 \\ 
0.5 \\
0.33
\end{pmatrix},\ 
\textbf{W}_{10,11} = \begin{pmatrix} 
0.6 \\
0.4
\end{pmatrix}.
$$

\begin{figure}[ht]
\centering
\includegraphics[width=0.4\textwidth]{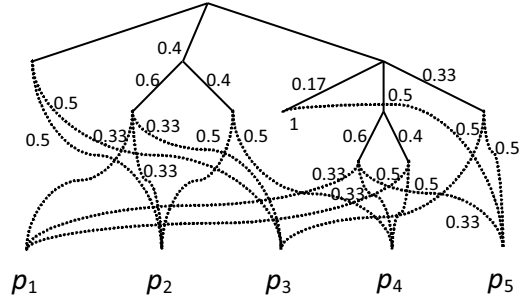}
\caption{Hierarchy for AHP}
\label{fig_3}
\end{figure}

For leaf nodes, we will index the weight coefficients by the number of the parent role and the number of permission assigned to the leaf: $w_{2,2} = w_{2,3} = 0.5$, $w_{5,1} = w_{5,2} = w_{5,3} \approx 0.33$, $w_{6,2} = w_{6,4} = 0.5$, $w_{7,5} = 1$, $w_{10,1} = w_{10,4} = w_{10,5} \approx 0.33$, $w_{11,1} = w_{11,4} = 0.5$, $w_{9,3} = w_{9,5} = 0.5$. In fig.~\ref{fig_3}, the leaves of the tree $T_p$ that correspond to the same permission are replaced by a single vertex, with the edges being assigned weights corresponding to their terminal vertices.

According to step~4 of the algorithm, the severity level of permission $p_1$ is calculated by the formula: 
$$S(p_1) = w_{5,1}w_5w_4 + w_{10,1}w_{10}w_8w_4 + w_{11,1}w_{11}w_8w_4 \approx$$ 
$$\approx 0.33 \cdot 0.6 \cdot 0.4 + 0.33 \cdot 0.6 \cdot 0.5 \cdot 0.4 + 0.5 \cdot 0.4 \cdot 0.5 \cdot 0.4 \approx 0.16.$$
Similarly:
$$S(p_2) \approx 0.5 \cdot 0.2 + 0.33 \cdot 0.6 \cdot 0.4 + 0.5 \cdot 0.4 \cdot 0.4 \approx 0.26.$$
$$S(p_3) \approx 0.5 \cdot 0.2 + 0.33 \cdot 0.6 \cdot 0.4 + 0.5 \cdot 0.33 \cdot 0.4 \approx 0.25.$$
$$S(p_4) \approx 0.5 \cdot 0.4 \cdot 0.4 + 0.33 \cdot 0.6 \cdot 0.5 \cdot 0.4 + 0.5 \cdot 0.4 \cdot 0.5 \cdot 0.4 \approx 0.16.$$
$$S p_5) \approx 1 \cdot 0.17 \cdot 0.4 + 0.33 \cdot 0.6 \cdot 0.5 \cdot 0.4 + 0.5 \cdot 0.33 \cdot 0.4 \approx 0.17.$$

Let us analyze the results obtained. As we can see from table~\ref{tab_1}, permissions $p_2$ and $p_3$ have the highest severity level, although their prevalence in the system is not maximal (Assumption~2). This is explained by the fact that the proposed approach takes into account also the location of the roles to which the interesting permission is assigned in the given hierarchy (Assumption~3). Indeed, the permissions $p_2$ and $p_3$ occur from the 2nd level of the role tree, whereas the most common in the system permissions $p_1$ and $p_4$ are present in the system, starting at level~3. Thus, in the example considered, the criterion of "proximity to the administrator" is more significant than the "prevalence in the system" criterion.

\begin{table}[ht]
\centering
\caption{Distribution of permissions in the role tree $T$}
\label{tab_1}
\begin{tabular}{|c|c|c|c|c|}
\hline
$p_i$    & Severity level & What roles are assigned & Number of roles & Levels of tree\\ 
\hline
\hline
$p_1$ & $0.16$ & $r_1$, $r_3$, $r_4$, $r_5$, $r_8$, $r_{10}$, $r_{11}$  & $7$ & $0$ -- $3$ \\
\hline
$p_2$ & $\textbf{0.26}$ & $r_1, r_2, r_3, r_5, r_6$ & $5$ & $0$ -- $2$ \\ 
\hline
$p_3$ & $\textbf{0.25}$ & $r_1, r_2, r_3, r_4, r_5, r_9$ & $6$ & $0$ -- $2$ \\
\hline
$p_4$ & $0.16$          & $r_1, r_3, r_4, r_6, r_8, r_{10}, r_{11}$ & $7$ & $0$ -- $3$ \\
\hline
$p_5$ & $0.17$          & $r_1, r_4, r_7, r_8, r_9, r_{10}$ & $6$ & $0$ -- $3$ \\
\hline
\end{tabular}
\end{table}

\section{Conclusions}

Most often, because there is no possibility to obtain quantitative estimates of the severity of certain processes, in practice a qualitative scale is used, for example: critical / high / medium / low. The article defines the algorithm for calculating the numerical indicators of the severity of the leakage of permissions in the role-based access control policy.

It follows from the proposed methodology that AHP makes it possible to automate the process of ranking permissions by significance or value from the point of view of information security. And the decision is made only on the basis of the features of the system itself, without involving the mechanism of expert assessments. This use of the method is new. Indeed, at present AHP is developing in the direction of using the theory of fuzzy sets, including within the framework of constructing a comprehensive information security system~\cite{b19,b20,b21,b22,b23}. But in the vast majority of works the method still relies on the mechanism of expert assessments.

The advantage and novelty of the proposed solution lies precisely in the "bundle" of AHP and the role tree of the access control system. This, on the one hand, allows you to obtain additional information in a situation where alternative solutions can not be linked by any precise functional dependencies, and on the other hand, relieves the method of inconsistency and subjectivism of expert judgments.

According to~\cite{b24}, the information security risk is an assessment of the possible damage to an organization or an asset as a result of a certain threat implementation. The main way to assess risks is to combine the probability of an event and its consequences:
$$R(V, T) = P(V, T) \cdot I(T),$$
where $P(V, T)$ is the probability of implementing the threat $T$ through a given vulnerability $V$ (for a two--factor risk assessment method) or the product of the probability of implementing the threat $T$ and exploiting vulnerability $V$ (for the three--factor risk assessment method), $I(T)$ is the damage from the implementation of the threat $T$~\cite{b25}. The main difficulty in solving the problems of quantitative assessment of information security risks is the qualitative nature of most indicators that affect the probability of the implementation of threats and the use of vulnerabilities, as well as determine the damage. The quantitative characteristics of permissions suggested in the article can be the basis for determining the probability of the implementation of information security threats through the vulnerabilities generated by the structure of roles of the access control policy.

%%%%%%%%%%%%%%%%%%%%%%%%%%%%%%%%%%%%%%%%%%%%%%%%%%%%%%%%%%%%%%%%%%%%%%%%%%%%%%%%%%%%%%%%%%%%%%

\end{document}